\documentclass[aps,prb,twocolumn,superscriptaddress,floatfix,amsmath,amssymb]{revtex4-2}
\usepackage{multirow}
\usepackage{graphicx}
\usepackage{bm}
\usepackage{dcolumn}
\usepackage{picture}
\usepackage[colorlinks=true,linkcolor=blue,urlcolor=blue,citecolor=blue]{hyperref}
\usepackage{natbib}
\usepackage{amsmath}
\usepackage{times}
\usepackage{color}
\usepackage{xcolor}
\usepackage{soul}
\usepackage{amssymb}

\begin{document}
\title{Sr$_3$LiIrO$_6$: a potential quantum spin liquid candidate in the one dimensional $d^4$ iridate family}
\author{Abhisek Bandyopadhyay}
\affiliation{School of Materials Sciences, Indian Association for the Cultivation of Science, 2A \& 2B Raja S. C. Mullick Road, Jadavpur, Kolkata 700 032, India}\affiliation{Department of Physics, Indian Institute of Science Education and Research, Pune, Maharashtra-411008, India}

\author{A. Chakraborty}
\affiliation{School of Physical Sciences, Indian Association for the Cultivation of Science, 2A \& 2B Raja S. C. Mullick Road, Jadavpur, Kolkata 700 032, India}\affiliation{Department of Physics, Indian Institute of Technology, Kanpur 208016, India}

\author{S. Bhowal}
\affiliation{School of Physical Sciences, Indian Association for the Cultivation of Science, 2A \& 2B Raja S. C. Mullick Road, Jadavpur, Kolkata 700 032, India}\affiliation{Materials Theory, ETH Zurich, Wolfgang-Pauli-Strasse 27, 8093 Zurich, Switzerland}

\author{Vinod Kumar}
\affiliation{Department of Physics, Indian Institute Of Technology Bombay, Powai, Mumbai 400076, India}

\author{M. M. Sala}
\affiliation{ESRF–The European Synchrotron, 71 Avenue des Martyrs, 38000 Grenoble, France}\affiliation{Dipartimento di Fisica, Politecnico di Milano, P.zza Leonardo da Vinci 32, I-20133 Milano, Italy}

\author{A. Efimenko}
\affiliation{ESRF–The European Synchrotron, 71 Avenue des Martyrs, 38000 Grenoble, France}
\author{C. Meneghini}
\affiliation{Dipartimento di Scienze, Universit\'{a} Roma Tre, Via della Vasca Navale, 84 I-00146 Roma, Italy}

\author{I. Dasgupta}
\affiliation{School of Physical Sciences, Indian Association for the Cultivation of Science, 2A \& 2B Raja S. C. Mullick Road, Jadavpur, Kolkata 700 032, India}

\author{T. Saha Dasgupta}
\affiliation{Department of Condensed Matter Physics and Material Sciences, S. N. Bose National Centre for Basic Sciences, Block JD, Sector 3, Saltlake, Kolkata -700106, India}

\author{A. V. Mahajan}
\affiliation{Department of Physics, Indian Institute Of Technology Bombay, Powai, Mumbai 400076, India}

\author{Sugata Ray}
\email[email:]{mssr@iacs.res.in}
\affiliation{School of Materials Sciences, Indian Association for the Cultivation of Science, 2A \& 2B Raja S. C. Mullick Road, Jadavpur, Kolkata 700 032, India}


\date{\today}

\begin{abstract}
Spin-orbit coupling (SOC) offers a large variety of novel and extraordinary magnetic and electronic properties in otherwise `ordinary pool' of heavy ion oxides. Here we present a detailed study on an apparently isolated hexagonal 2$H$ spin-chain $d^4$ iridate Sr$_3$LiIrO$_6$ (SLIO) with geometric frustration. Our structural studies clearly reveal perfect Li-Ir chemical order in this compound. Our combined experimental and {\it ab-initio} electronic structure investigations establish a magnetic ground state with finite Ir$^{5+}$ magnetic moments in this compound, contrary to the anticipated nonmagnetic $J$=0 state. Furthermore, the dc magnetic susceptibility ($\chi$), heat capacity ($C_p$) and spin-polarized density functional theory (DFT) studies unravel that despite having noticeable antiferromagnetic correlation among the Ir$^{5+}$ local moments, this SLIO system evades any kind of magnetic ordering down to at least 2 K due to geometrical frustration, arising from the comparable interchain Ir-O-O-Ir superexchange interaction strengths, hence promoting SLIO as a potential quantum spin liquid candidate. 
\end{abstract}

\maketitle


\newpage

\section{Introduction}
Magnetic frustration disrupts conventional long-range possible magnetic order and establishes a highly entangled ground state with nonlocal excitations known as quantum spin liquid (QSL)~\cite{1,3,4,5,6}. 
Materials supporting strongly quantum-entangled spin-liquid states are potential for data storage and memory applications, and in particular, it is possible to realize topological quantum computation by means of QSL states. 

In this context, 
it is of interest to explore the enhanced quantum fluctuations of the spin-orbital entangled states of the geometrically frustrated 5$d$ Ir-oxides. 
Here one of the most discussed prevailing controversy in the context of pentavalent $d^4$ (Ir$^{5+}$:5$d^4$) iridates is the deviation of ideally nonmagnetic $J$ = 0 state under strong SOC limit, and the concomitant development of finite Ir-magnetic moments~\cite{a2sciro6-inorgchem,doped-sriro3,nagprl1,nagprb-6H,ownprb-psmio,antisite5+prb} and QSL in these systems.
\par
In such a backdrop, the 2$H$-prototype $A_3$$A^{\prime}$$M$O$_6$ ($A$ = Sr, Ca; $A^{\prime}$, any transition metal or nonmagnetic cation; $M$ = transition metal) family of hexagonal oxides appear interesting because of their unconventional magnetic properties due to the interplay between low dimensionality, magnetic frustration, and magnetocrystalline anisotropy~\cite{sr3niiro6-prb-dta}. More recently the exploration of this family of compounds has been extended towards the analogs containing heavier 4$d$/5$d$ transition-metal ions, with the aim of studying the novel ground state magnetic behaviors in the strong spin-orbit coupling regime~\cite{sr3niiro6-prb-dta,ca3corho6-jssc,sr3coiro6-prb,sr3nirho6-prb,sr3niiro6-ej}. Due to spatially extended 5$d$ electronic orbitals, the extended superexchange interactions (via multiple oxygen ions) between the neighboring spin-chains effectively influence their respective magnetic ground states~\cite{jacs132,inorgchem50,prb86,sr3naruo6-prb}. It would therefore be an interesting proposition to check if, by introducing Ir$^{5+}$ at the $M$-site and any nonmagnetic monovalent alkali metal cation at the $A^{\prime}$-site, the strongly spin-orbit coupled Ir$^{5+}$-magnetism can be elucidated in the limit of geometrical frustration and a path for realizing the quantum-entangled spin-liquid phase could be evidenced in such an apparently reduced dimensional structure.
\par
Here in this paper we present detailed structural, chemical, electronic, magnetic and thermodynamic characterizations of a $d^4$ columnar spin-chain iridate Sr$_3$LiIrO$_6$ (SLIO). By means of systematic investigations on the structural and chemical properties we infer that this system is absolutely free from structural disorder between the Li- and Ir-sites as well as Sr- and Li-sites. The x-ray absorption (XAS) and x-ray photoemission spectroscopy (XPS) studies confirm pure single 5+ oxidation state of Ir in this SLIO compound. The bulk magnetization study together with the spin-polarized electronic structure calculations strongly suggest that each Ir$^{5+}$ ion in SLIO possesses a large magnetic moment ($\sim$0.45 $\mu_B$), thus refuting strong spin-orbit coupled atomic nonmagnetic $J$=0 picture for Ir$^{5+}$ in SLIO. Furthermore, the detailed dc magnetic susceptibility ($\chi$) and heat capacity ($C_p$) measurements, as well as theoretical characterizations unveil that despite having significant antiferromagnetic nearest-neighbor interaction ($\Theta_{CW}$ $\approx$-71 K) between the large local Ir$^{5+}$ moments, the SLIO does not magnetically order down to at least 2 K (frustration parameter $>$35) due to geometric frustration, arising in the edge-shared Ir-triangular network (Fig. 5(b)) as a result of the comparable interchain Ir-Ir magnetic exchange interaction strengths, thereby pushing SLIO towards a potential QSL state. In addition, the magnetic heat capacity displays a linear temperature-dependence at low temperatures, a quintessential feature for metals but here in a spin-orbit Mott insulator, pointing towards the gapless nature of spin excitations within this compound. 
\section{Methodology}
\subsection{Experimental techniques}
Polycrystalline Sr$_3$LiIrO$_6$ sample was prepared using conventional solid state reaction technique from high purity ($>$ 99.9\%) SrCO$_3$, Li$_2$CO$_3$ and IrO$_2$ powders. The stoichiometric amounts of all these starting materials were mixed and homogeneously ground in an agate mortar. The mixture was initially calcined at 600$^{\circ}$C in air for 10 hours. The resultant mixture was then reground and re-annealed at several higher temperatures (700$^{\circ}$C and 800$^{\circ}$C for 12 hours each) in air with few intermediate grindings. The phase purity of the sample was checked from X-ray powder diffraction (XRD) measured at Bruker AXS: D8 Advance x-ray diffractometer with Cu K$_{\alpha}$ radiation at the room temperature. The sample's structure was obtained after analyzing the XRD data by the Rietveld technique using FULLPROF program~\cite{fullprof}. To verify chemical homogeneity and cation-stoichiometry in the sample, energy dispersive x-ray (EDX) analysis was also performed using a field emission scanning electron microscope (JEOL, JSM-7500F). The cation-stoichiometry was also checked by inductively coupled plasma-optical emission spectroscopy (ICP-OES) using a Perkin Elmer Optima 2100 DV instrument. The Ir $L_3$-edge X-ray Absorption Near Edge structure (XANES) and Extended X-ray Absorption Fine Structure (EXAFS) measurements have been performed at the XAFS beamline of Elettra (Trieste, Italy) synchrotron radiation facility at room temperature in standard transmission geometry. Data treatment and quantitative analysis of the measured EXAFS data have been carried out using ARTEMIS program~\cite{artemis,ravel}. The core level and valence band x-ray photoemission spectroscopy (XPS) measurements were carried out using an Omicron electron spectrometer, equipped with a Scienta Omicron sphera analyzer and an Al $K_{\alpha}$ monochromatic source with an energy resolution of 0.5 eV. The sample surface was cleaned before experiment by {\it in situ} Ar sputtering to negate the surface oxidation effect as well as the presence of environmental carbons in the pelletized sample. The collected XPS spectra were processed and analyzed with the KOLXPD program. Further, the electrical resistivity was measured using standard four-probe method within the temperature range of 100-400 K in a lab-based resistivity setup. The $dc$ magnetic susceptibility was measured in the temperature range of 2$-$400 K and in magnetic fields up to $\pm$70 kOe in a superconducting quantum interference device (SQUID) magnetometer, Quantum Design. Further, heat capacity in both zero field and applied magnetic fields was measured within 2$-$300 K temperature range in a physical property measurement system (PPMS, Quantum Design). Moreover, the Ir $L_3$-edge resonant inelastic x-ray scattering (RIXS) was measured on this sample at the ID20 beamline of the European Synchrotron Radiation Facility (ESRF) using $\pi$-polarized photons and a scattering geometry with 2$\theta$ $\simeq$ 90$^{\circ}$ to suppress elastic scattering.
\subsection{Theoretical methods}
The electronic structure calculations based on density functional theory (DFT) presented in this paper are carried out in the plane-wave basis within generalized gradient approximation (GGA)\cite{gga} of the Perdew-Burke-Ernzerhof exchange correlation supplemented with Hubbard U as encoded in the Vienna \textit{ab-initio} simulation package (VASP) \cite{vasp1,vasp2} with projector augmented wave potentials \cite{aug1,aug2}. The calculations are done with usual value of U and Hund’s coupling (J$_H$) chosen for Ir with U$_{eff}$($\equiv$U-J$_H$) = 1.5 eV \cite{refu1,arxiv-snio} in the Dudarev scheme \cite{dudarev}. In order to achieve convergence of energy eigenvalues, the kinetic energy cut off of the plane wave basis is chosen to be 550 eV. The Brillouin-Zone integrations are performed with 8$\times$8$\times$8 Monkhorst grid of k-points. The symmetry protected ionic relaxation of the experimentally obtained crystal structure has been carried out within VASP calculation using the conjugate-gradient algorithm until the Hellman-Feynman forces on each atom were less than the tolerance value of 0.01 eV/\AA. The optimized structure has been used for calculations.
\par To get a quantitative estimate of the noncubic crystal distortion mediated Ir-t$_{2g}$ crystal field splitting and realistic hopping parameters of our low-energy Hamiltonian, we employed the muffin-tin orbital (MTO) based Nth order MTO (NMTO) method\cite{nmto1,nmto2}, as implemented in Stuttgart code, retaining Ir-t$_{2g}$ orbitals within the basis set and down-folding higher degrees of freedom. The NMTO method relies on the self-consistent potentials borrowed from the linear MTO (LMTO) calculations\cite{lmto}. For the self-consistent LMTO calculations within the atomic sphere approximation (ASA), the space filling in the ASA is obtained by inserting appropriate empty spheres in the interstitial regions.

\section{Results and Discussion}
\subsection{Structure from x-ray diffraction and composition verification}
Rietveld refinement of the powder x-ray diffraction (PXRD) pattern [Fig. 1(a)], collected from the SLIO sample at 300 K, confirms pure single phase with rhombohedral $R$$\bar{3}$c space group. The refined crystal structure [see Fig. 1(b)] consists of alternately arranged face-shared LiO$_6$ trigonal prisms and IrO$_6$ octahedral units, leading to the infinite one dimensional (1D) chains along the $c$-axis. These chains are separated from each other by nonmagnetic Sr cations. In addition, presence of any anti-site disorder between Li and Ir has been clearly refuted from our structural refinement. The refined structural parameters are indicated in Table-I. Further, the possibility of any site-exchange between Sr (18$e$) and Li (6$a$) has also been discarded from the XRD refinement, supporting the previously reported~\cite{sliojmc} 2$H$-prototype columnar spin-chain structure of this SLIO compound. In addition, the IrO$_6$ octahedra undergo weak trigonal distortion in terms of the little deviation of O-Ir-O bond angles ($\sim$ 90.4$^{\circ}$) from the ideal 90$^{\circ}$ of perfectly cubic.
\begin{figure}
\begin{center}
\resizebox{8.6cm}{!}
{\includegraphics[15pt,476pt][561pt,701pt]{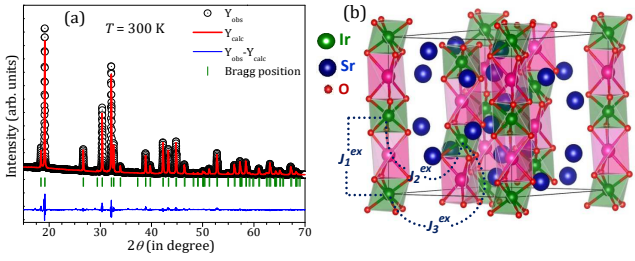}}
\caption{(a) Rietveld refined XRD pattern of Sr$_3$LiIrO$_6$. (b) The refined crystal structure showing three different exchange interactions by blue dotted lines/curves.}
\end{center}
\end{figure}
\begin{table}[h]
\begin{center}
\caption{\it Structural information extracted from Rietveld refinement of XRD data of Sr$_3$LiIrO$_6$ sample at 300 K. Space group: $R$$\bar{3}$c, $a$ = $b$ = 9.6429(1) {\AA}, $c$ = 11.1442(3) {\AA}, $\gamma$ = 120$^{\circ}$, $V$ = 899.5146(5) {\AA}$^3$}
\resizebox{8.6cm}{!}{
\begin{tabular}{| c | c | c | c | c | c |}
\hline Atoms & site & occupancy &  $x$ & $y$ & $z$ \\\hline
  Sr & 18e & 1.0 & 0.3579(7) & 0 & 0.25 \\
  Li & 6a & 1.0 & 0 & 0 & 0.25 \\
  Ir & 6b & 1.0 & 0 & 0 & 0 \\
  O & 36f & 1.0 & 0.1757(4) & 0.0212(5) & 0.1071(6) \\
\hline
\end{tabular}}
\end{center}
\end{table}
The stoichiometry of this sample has been verified by SEM-EDX, which confirmed chemical homogeneity of SLIO. In addition, the cation stoichiometry is almost retained at the target composition, Sr : Na : Ir = 2.998:0.998:1.00 $\equiv$ 3 : 1 : 1 within the given accuracy of the measurement. The cation-stoichiometry was further quantified through ICP-OES analysis which also reveals the actual stoichiometry to be at the desired level.
\begin{figure}
\begin{center}
\resizebox{8.6cm}{!}
{\includegraphics[9pt,56pt][590pt,765pt]{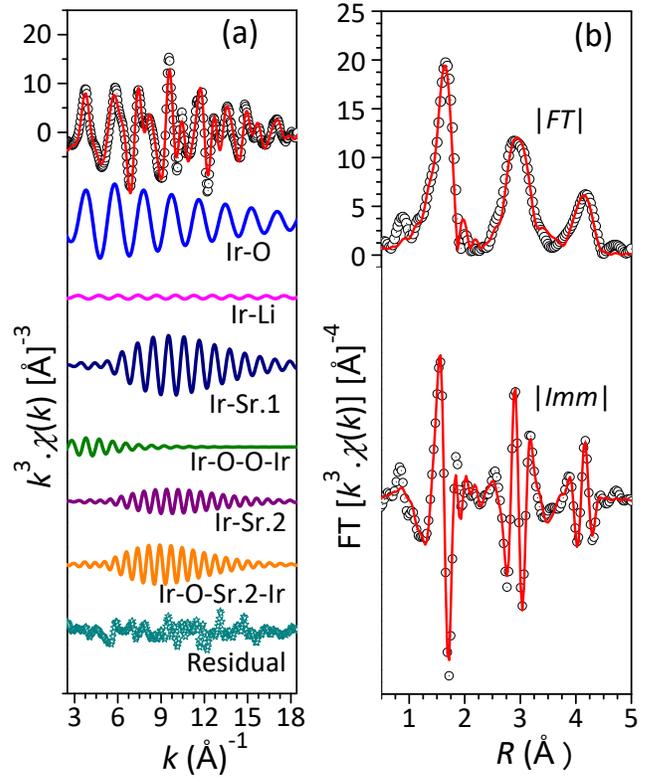}}
\caption{(Left Panel) $k$$^3$ weighted Ir $L_3$-edge experimental EXAFS data (open black circles) and the respective best fit (red solid line) for Sr$_3$LiIrO$_6$ sample (a). The contributions from the individual single  and multiple scattering paths (solid colored line) and the residual [$k^3$$\chi$$_{exp}$-$k^3$$\chi$$_{th}$] (open cyan stars) are also indicated, vertically shifted for clarity. (Right Panel) Fourier transform of the respective experimental (open black circles) and theoretical (solid red line) curves (b); the magnitude ($|$FT$|$) and the imaginary parts ($Imm$) are also indicated; vertically shifted for clarity.}
\end{center}
\end{figure}
\subsection{Local structure from Ir $L_3$-edge extended x-ray absorption fine structure (EXAFS)}
It is known that the microscopic detail of any system may take up a very different arrangement at the local level keeping a deceptive similarity in the global context~\cite{vmo-ownprb,lsvmoprb}. On the other hand, local anti-site disorder in 5$d$ iridates largely influences the ground state magnetic responses~\cite{byio-asd}. The same is also pertinent in case of SLIO, as the Li/Ir antisite disorder would eventually increase the face-sharing Ir-Ir connectivity in this system. Therefore, in order to check the local atomic distribution along the 1D chain of this compound, the local structure has been investigated by measuring the Ir $L_3$-edge EXAFS signal. The collected experimental EXAFS data along with the theoretical fitting are represented in Fig. 2 (a) and (b). The fitted structural parameters obtained from EXAFS data analysis are tabulated in Table-II. The multishell data refinement procedure~\cite{vmo-ownprb} clearly support the hypothesis of full Li/Ir site-ordering in this sample, consistent with the XRD refinement. Moreover, the local interatomic distances obtained from the EXAFS fitting are also consistent with those found from XRD fiting (see Table-II).
\begin{table}[h]
\begin{center}
\caption{The results obtained from the Multi-shell analysis of the Ir $L_3$-edge EXAFS spectrum. Constraints among the parameters were applied in order to reduce the correlation among them. The fixed or constrained values are labeled by `$\ast$'. The absolute mismatch between the experimental data and the best fit is $R$$^{2}$ = 0.02. The bond distances, obtained from  XRD refinement, are also indicated for the sake of comparison.}
\resizebox{8.6cm}{!}{
\begin{tabular}{| c | c | c | c | c |}
\hline Shell & $N$ & $\sigma$$^{2}$ ($\times$ 10$^{2}${\AA}$^2$) & $R$({\AA}) & $R$$_{XRD}$({\AA}) \\\hline
   Ir-O & 6.0$^{\ast}$ & 0.237(1) & 1.96(2) & 1.96(2)\\
   Ir-Li & 2.0 & 0.078(3) & 2.701(2) & 2.786(1)\\
   (intrachain) &  &  &  &  \\
   Ir-Sr1 & 6.0$^{\ast}$ & 0.52(3) & 3.234(2) & 3.236(5)\\
   Ir-O-O (MS) & 6.0$^{\ast}$ & 0.439(2) & 4.01(5) & 4.26(5)\\
   Ir-Sr2 & 6.0$^{\ast}$ & 0.61(3) & 4.29(2) & 4.439(2)\\
\hline
\end{tabular}}
\end{center}
\end{table}
\begin{figure}
\resizebox{8.6cm}{!}
{\includegraphics[35pt,510pt][525pt,728pt]{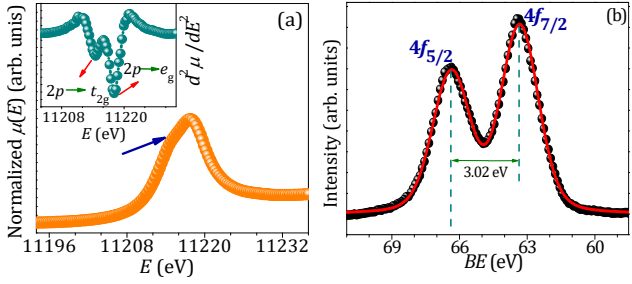}}
\caption{(a) Ir-$L_3$ edge XANES spectrum (shaded orange circles), with the solid blue arrow indicating asymmetric shoulder; Inset: corresponding second derivative curve (shaded dark cyan circles); (b) Ir 4$f$ core level XPS spectra (shaded black circles) along with the fitting (red solid line).}
\end{figure}
\subsection{Electronic characterization}
Estimation of Ir-oxidation state has the central importance in the context of ground state magnetism of 5$d$ iridates, as Ir$^{4+}$ and Ir$^{6+}$ species are strongly magnetic~\cite{sioprl,lzioprb,sriro3prb,sziojssc}, while the Ir$^{5+}$ ion is ideally nonmagnetic $J$ = 0~\cite{6H10}. On the other hand, deviation from the ideal nonmagnetic $J$ = 0 state in pentavalent $d^4$ iridates is often attributed to the presence of magnetic Ir$^{4+}$/Ir$^{6+}$ impurities~\cite{6H13}. Therefore, to confirm the Ir-valence in SLIO, the Ir $L_3$-edge x-ray absorption near edge structure (XANES) spectrum has been collected and shown in Fig. 3(a). The asymmetric structure of the spectral line, in the form of a weak shoulder in the lower energy side (shown by the solid blue line in Fig. 3 (a)), resembles quite well with the higher oxidation state ($>$ 4+) of Ir~\cite{irxanes1,irxanes2}. The corresponding second derivative curve (inset to Fig. 3 (a)), representative of the white-line feature, clearly reveals well-resolved doublet feature, identifying the 2$p$ $\rightarrow$ $t_{\text{2}g}$ (lower energy feature) and 2$p$ $\rightarrow$ $e_g$ (higher energy peak) transitions. The peak shape, structure, and the relative peak intensity corresponding to 2$p$ $\rightarrow$ $t_{\text{2}g}$ transition (inset to Fig. 3(a)) strongly affirm pure 5+ oxidation state of Ir~\cite{ownprb-psmio,irxanes1,irxanes2} in SLIO.

The Ir 4$f$ core level XPS spectrum was measured and subsequently fitted using a single spin-orbit split doublet [Fig. 3(b)]. The energy positions of the 4$f_{7/2}$ and 4$f_{5/2}$ doublets along with their spin-orbit separation of 3.02 eV further confirm pure 5+ valence of Ir in this compound~\cite{nagprl1,nagprb-6H,ownprb-psmio,nagprbdp,bcio-ownprb}.

\subsection{Electrical resistivity, XPS valance band, non-spin polarized electronic structure and Ir-$L_3$ resonant inelastic x-ray scattering (RIXS) studies}
Similar to the observation of spin-orbit coupled several 5$d$ iridates~\cite{nagprl1,nagprbdp,bcio-ownprb,lszioprb}, the negative temperature dependence of the measured electrical resistivity of SLIO (see Fig. 4(a)) suggests insulating behavior in the entire temperature range of measurement. In addition, XPS valance band spectrum (inset to Fig. 4(a)) clearly reveals absence of finite density of states at the Fermi level (shown by vertical green dashed line in the inset of Fig. 4(a)), affirming the charge-gapped electronic state for SLIO. Such a gapped electronic behavior establishes importance of SOC in Sr$_3$LiIrO$_6$ too.

\begin{figure}
\resizebox{8.6cm}{!}
{\includegraphics[0pt,0pt][463pt,375pt]{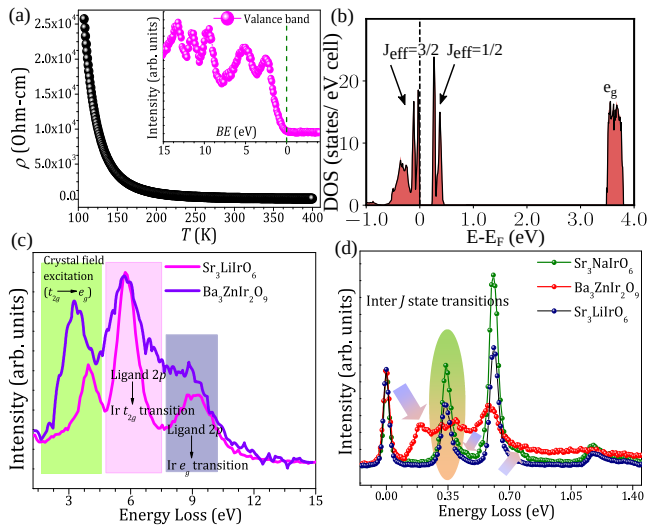}}
\caption{(a) Temperature dependence of electrical resistivity (shaded black circles); Inset: XPS valance band spectrum (shaded violet circles), with the vertical green dashed line representing the Fermi energy position $E_F$; (b) The orbital resolved density of states of Ir-d orbital for SLIO within GGA+SOC; Comparison of the Ir-$L_3$ RIXS spectra in (c) low-resolution high-energy features and (d) high-resolution low-energy excitations.}
\end{figure}
 At this point, we have carried out non-spin polarized GGA+SOC electronic structure calculations of Sr$_3$LiIrO$_6$ (SLIO). The GGA + SOC  Ir-d projected  density of states plot in Fig.4 (b) reveal strong crystal field splitting of $\sim$3.9 eV within $t_{2g}$ and $e_g$ of Ir-d states due to the local octahedral environment of the Ir atoms, which is consistent with the low-resolution high energy RIXS results. Presence of trigonal distortion within the IrO$_6$ octahedra further lifts the degeneracy of the $t_{2g}$ states into low lying singly degenerate $a_{1g}$ and higher lying doubly degenerate $e^\pi_g$ states with small non-cubic crystal field splitting of value 0.03 eV.  Due to presence of heavy Ir atoms, spin orbit coupling (SOC) has profound impact on the electronics structure of the SLIO. Inclusion of SOC breaks the Ir-$t_{2g}$ states into $J_{eff}= 3/2$ ($\Gamma_8$) and $J_{eff}={1/2}$ ($\Gamma_7$) states as seen in the density of states plot of Fig.4 (b). The four electrons of the valence Ir-d orbital completely occupy the pseudo-spin $J_{eff}= 3/2$ states keeping the $J_{eff}={1/2}$ state completely empty.
 \par The near isolation of Ir atoms in SLIO is supported by the presence of narrow crystal field excitation ($\sim$3.9 eV) and charge transfer ($\sim$6 and 9 eV) peaks in the low-resolution Ir-$L_3$ RIXS spectrum of SLIO compared to the 6$H$-hexagonal Ba$_3$ZnIr$_2$O$_9$~\cite{nagprl2} (see Fig. 4(c)).
 The band-width of $t_{2g}$ states for SLIO is found to be $\sim$0.8 eV which is slightly higher than that of it's sister compound Sr$_3$NaIrO$_6$ (band-width $\sim$0.7 eV)\cite{arxiv-snio}. but substantially smaller than Ba$_3$ZnIr$_2$O$_9$ \cite{nagprl2}.
 Another notable feature to address is that the $t_{2g}$ - $e_g$ crystal field splitting energy for SLIO ($\sim$3.9 eV) in the present study appears at relatively higher energy compared to that of SNIO ($\sim$3.5-3.6 eV)~\cite{arxiv-snio} and 6$H$-BZIO ($\sim$3.2-3.3 eV)~\cite{nagprl2}, illustrated in Fig. 4(c), which could be assigned to the shorter Ir-O bond lengths (1.96 \AA) in SLIO than those of the SNIO and BZIO ($\approx$2.03 and 1.9-2.0 {\AA} respectively) \cite{nagprl1,arxiv-snio}, and the lower level of Ir-$t_{2g}$ splitting due to much lesser IrO$_6$ octahedral distortion in SLIO with respect to the SNIO and BZIO \cite{nagprl1,arxiv-snio}.
So, it would now be quite natural to assume that a combination of lesser effective spatial dimensionality ({\it i.e.,} isolated nature Ir$^{5+}$ ions), comparatively smaller bandwidth and dominance of spin-orbit coupling in SLIO could extend a much higher possibility to realize the coveted $J_{eff}$=0 nonmagnetic ground state within the nearly atomic SOC limit.

Next we compare the low-energy, high-resolution Ir-$L_3$ edge RIXS spectra between the two isovalent isostructural columnar iridates Sr$_3$(Na,Li)IrO$_6$ and a 6$H$ hexagonal iridate Ba$_3$ZnIr$_2$O$_9$ (BZIO), measured within the same technical specifications \cite{nagprl2}, and the results are resumed in Fig. 4(d). It is evident that there are substantial differences in the spectral width, intensity, energy positions as well as number of spectral features (peak at 0.18 eV energy loss and double peak feature in the 0.3-0.4 eV range of the energy loss in BZIO, shown by the bicoloured arrow and ellipse respectively, which are missing in both the columnar systems) between the columnar and 6$H$ family of $d^4$ iridates \cite{nagprl2,arxiv-snio} could certainly be ascribed to the enhanced degree of intersite Ir-Ir hopping due to face-sharing IrO$_6$ octahedral connectivity, as well as greater extent of Ir $t_{2g}$ trigonal crystal field splitting in BZIO \cite{nagprl1} compared to the columnar Sr$_3$NaIrO$_6$ \cite{arxiv-snio} and Sr$_3$LiIrO$_6$ systems. On the other hand, like in Sr$_3$NaIrO$_6$ (SNIO) \cite{arxiv-snio}, we observe three similar inelastic peaks below 1.5 eV in SLIO. Although the shape and energy positions of these three peaks overall appear similar in both the columnar iridates SNIO and SLIO, subtle changes in these inelastic RIXS features are clearly visible, as demonstrated by intensity enhancement and slight peak width broadening, as well as development of weak but noticeable shoulders in the higher-energy sides of the respective first and second inelastic peaks (marked by two bicoloured arrows for SNIO in Fig. 4(d)). Clearly these differences in spectral features between the two columnar iridates lie in their respective Ir-O octahedral distortions, Ir-Ir hopping pathways and extent of Na/Li-Ir anti-site disorder. While both these columnar iridates generally accommodate common face-sharing geometry along $c$-axis of the crystal structure, the SLIO possesses comparatively weaker extent of trigonal distortion in terms of less deviated O-Ir-O bond angles (90.4$^{\circ}$) from ideal 90$^{\circ}$ with respect to SNIO (91.68$^{\circ}$), and at the same time, there is increased number of direct face-sharing Ir-Ir hopping pathways along the chain in anti-site disordered SNIO ($\sim$10\% Na-Ir disordering)~\cite{arxiv-snio} compared to the nominal Ir-O-(Li)-O-Ir intrachain extended superexchange pathways in case of perfectly structurally ordered SLIO. Also, there are different types of interchain Ir-Ir connectivity and the respective hopping networks (via Ir-O-O-Ir) as a consequence of Na-Ir site-disorder in addition to the nominally Na-Ir ordered configurations in SNIO, in contrary to only two Ir-Ir interchain distances (schematically presented in Fig. 1(b) by $J_2^{ex}$ and $J_3^{ex}$ at $\sim$5.87 and 6.7 \AA respectively) and the corresponding Ir-O-O-Ir hopping pathways of the full Li-Ir chemically ordered SLIO. Moreover, there would be bimodal distributions of local noncubic crystal fields, Ir-O covalent interactions, Ir-Ir superexchange interactions $J_{SE}$($\equiv$4$t^2$/$U$) and the effective SOC strengths in SNIO due to the presence of $\sim$90\% Ir in the ordered sites along with the $\sim$10\% Ir in the antisite positions \cite{arxiv-snio}, which is naturally absent in the complete site-ordered SLIO. All these issues give rise to different extent of lifting of Ir-$t_{2g}$ degeneracy, and accordingly, rearranging the spin-orbit derived Ir-$J_{eff}$ states in either of these samples. In such a scenario, we must infer that the aforementioned changes in low-energy Ir-$L_3$ RIXS features of Sr$_3$LiIrO$_6$ relative to the Sr$_3$NaIrO$_6$ and Ba$_3$ZnIr$_2$O$_9$ should be due to the dissimilar Ir-Ir hopping connectivity and also the influence of non-identical trigonal crystal field splitting on the Ir $t_{2g}$ orbitals in this SLIO. Infact, both hopping and noncubic crystal field effect impart a strong impact on the effective SOC strength in $d^4$ iridate systems~\cite{nagprl2}. Therefore, demonstrating the spin-orbit-coupled Ir energy levels from the perspective of atomic $J$ picture solely, as has been the widely accepted description until recently \cite{psmio-49}, becomes insufficient, as
revealed by Nag {\it et al.}~\cite{nagprl2} and Revelli {\it et al.}~\cite{psmio-50} very recently. So, as estimation of the effective spin-orbit coupling strength $\lambda_{eff}$ on Ir within only a purely atomic limit is not at all a reasonable approach, we refrain from doing such an estimation here in SLIO. 

\begin{figure*}
\resizebox{14.5cm}{!}
{\includegraphics[19pt,400pt][575pt,759pt]{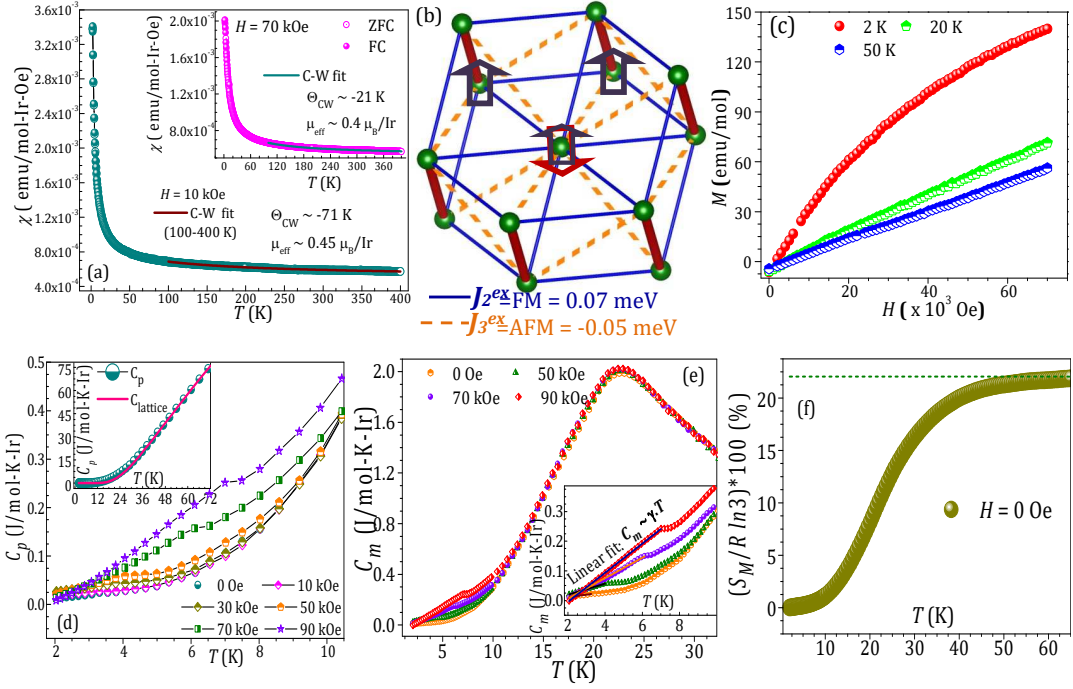}}
\caption{(a) Temperature dependence of dc magnetic susceptibility data along with Curie-Weiss fitting for 10 kOe (main panel) and 70 kOe (inset) fields; (b) The edge-shared frustrated Ir-triangular network formed by interchain Ir-Ir exchanges $J_2^{ex}$ and $J_3^{ex}$; (c) Field dependent isothermal dc magnetization $M-H$ variations at three different temperatures;(d) Temperature variations of total specific heat $C_p$ in the low-$T$ region for zero field and applied magnetic fields, Inset: Lattice part extraction (solid pink line) from the zero field $C_p$ data (half-filled cyan circles); Temperature dependence of (e) magnetic specific heat $C_m$, and (f) Zero field magnetic entropy $S_M$.}
\end{figure*}

\subsection{dc magnetic susceptibility and spin polarized DFT calculations}
Now in order to unfold the crucial question of whether the elusive $J_{eff}$ = 0 state is realized in SLIO, or not, the temperature dependences of the zero-field-cooled (ZFC) and field-cooled (FC) {\it dc} magnetic susceptibility have been measured under different applied magnetic fields, and the results are summarized in Fig. 5(a). The featureless paramagnet-like bulk susceptibility clearly refutes any kind of magnetic ordering down to 2 K. However, for low applied fields ($<$5000 Oe, not shown in the figure), appearance of a noticeable thermo-magnetic irreversibility between the ZFC and FC magnetizations from a higher temperature, and a gradual suppression of this low-field-ZFC/FC-divergence upon increasing applied magnetic fields, indicate presence of short-range magnetic correlations and/or minor spin-freezing effect in this sample~\cite{4,chap7-19,chap7-20}. Further, the field-dependent isothermal magnetization $M-H$ curves, as presented in Fig. 5(c), clearly reveal absence of coercivity and remanent magnetization even at the lowest measured temperature of 2 K, but the slight nonlinearity probably supports the development of spin correlations in SLIO at low temperatures, quite common in $d^4$ iridates~\cite{6H13,cdir2-12,cdir2-38}.

Now in such disordered magnetic system, it should be worthwhile to state that within weak temperature-dependence of the measured dc susceptibility, the Curie-Weiss (C-W) fitting parameters always carry some uncertainties which critically depend on both the temperature range of C-W fitting and the applied magnetic fields of measurement~\cite{nagjmmm}. Considering all the possibilities, the C-W analysis ($\chi$ = $\chi_0$ + $\frac{C_W}{(T - \Theta_{CW})}$ with $\chi_0$ being the temperature independent paramagnetic susceptibility, and $C_W$, $\Theta_{CW}$, the Curie constant and the Curie-Weiss temperature, respectively) on the 10 kOe field-cooled $\chi$($T$) data seems to be most reasonable in the temperature range 100-400 K (see Fig. 5(a)). The resulting fit yields a $\Theta_{CW}$ value $\approx$ -71 K and an effective paramagnetic moment $\mu_{eff}$ $\sim$ 0.45 $\mu_B$/Ir.
\par To explore the possibility of magnetism, the spin polarized GGA+SOC+U calculation keeping the spin quantization axis along [001] direction, has been carried out of the antiferromagnetic (AFM) configuration having anti-parallel spins orientations at the neighboring Ir sites within the linear chain. Total energy calculated from spin-polarized density functional theory in presence of SOC suggests AFM configuration to be lower in energy than both of the non-magnetic and ferromagnetic configuration by energy values $\Delta E$/f.u.$\sim$ 2~meV and 3~meV respectively. This suggests the presence of AFM interactions in SLIO, consistent with the negative $\Theta_{CW}$ extracted from the high temperature Curie-Weiss fit of the dc susceptibility measurements. In summary, our calculations reveal a magnetic ground state with AFM spin configurations and large Ir$^{5+}$ magnetic moments in contrast to the non-magnetic $J_{eff}=0$ state, as suggested by Ming \textit{et. al.} in a recent theoretical study \cite{ming}.\\
Moreover, the intra and inter-chain hopping interaction strengths between the Ir-$d$ orbitals are calculated to be $\sim$ 130 meV and $\sim$ 76 meV respectively. The estimated substantial nearest neighbor inter-chain Ir-Ir hopping interaction through inter-orbital Ir-d$_{xz}$ and Ir-d$_{yz}$ overlap suggests three-dimensional nature of magnetic exchange within this apparently one-dimensional structure of SLIO. This observation of significantly large magnetic moments on individual Ir$^{5+}$ ion immediately refutes the nonmagnetic $J_{eff}$=0 ground state proposition, and thus, an isolated atom-like description of Iridium in SLIO, and establishes SLIO as a magnetically 3-dimensional (3D) system, consistent with the previously reported isostructural columnar spin-chain compounds \cite{sr3naruo6-prb,snio-4,snio-6,snio-9}. Further, this large value of Ir$^{5+}$ magnetic moment strongly endorses only an intermediate effective SOC picture~\cite{snio-8,cdir2-18} in SLIO, thereby opposing the anticipated higher atomic Ir-SOC picture in this compound.\\
The origin of such a large Ir$^{5+}$-moment in SLIO could certainly be attributed to the intersite Ir-Ir hopping via extended Ir-O-O-Ir superexchange pathways within all possible intra and interchain magnetic exchange connectivity (represented by three different exchange interactions in Fig. 1 (b)), causing delocalization of intra-site Ir$^{5+}$ holes, and therefore, creating deviation from a perfect atomic $J$ = 0 arrangement~\cite{ownprb-psmio,nagprbdp,bcio-ownprb}. In addition, the trigonal distortion of the IrO$_6$ octahedra and the subsequent Ir-$t_{2g}$ splitting, as well as the hybridization of extended Ir-5$d$ orbitals and the stronger Ir-O covalent interactions together might cause rearrangement in the spin-orbit coupled Ir energy levels in SLIO, and as a result, leading to the possible breakdown of atomic $J$ limit~\cite{cdir2-18,cdir2-17,Takegami-prb2020,Paramekanti-prb2018}. But surprisingly, despite having significantly strong antiferromagnetic correlation (by means of large negative $\Theta_{CW}$) between the large local Ir$^{5+}$ moments, the SLIO compound does not trigger magnetic ordering down to 2 K at least.

To corroborate our prediction of magnetism within SLIO, we have calculated the magnetic exchange interactions between the Ir atoms. For a quantitative estimation of the Ir-Ir exchange couplings, we have calculated the symmetric exchange interactions by mapping total energies of several spin configurations within GGA+SOC+U scheme to the Heisenberg model $H_{spin}= \sum_{ij}J_{ij}{\vec S}_i \cdot {\vec S}_j$. Our calculations show that the nearest neighbor intra-chain Ir-Ir interaction $J_1^{ex}$ is dominant and AFM in nature with a magnitude of 4.26 meV. The second ($J_2^{ex}$) and third ($J_3^{ex}$) nearest-neighbor interactions, representing the inter-chain Ir-Ir exchanges, are relatively weaker with magnitude of 0.07 meV and 0.05 meV and respectively FM and AFM in nature. The twelve inter-chain $J_2^{ex}$ and $J_3^{ex}$ bonds with comparable exchange interaction strengths therefore form a frustrated edge-sharing triangular network (see Fig. 5b), prohibiting the long range ordering within this system.

Finally, absence of magnetic order within persistent significant antiferromagnetic (AFM) correlation (in the form of large negative $\theta_{CW}$ value) is one of the signature characteristics of quantum spin-liquids (QSLs). From this view point therefore, the Sr$_3$LiIrO$_6$ compound could be proposed as a potential QSL candidate. Further, certain ambiguity in the AFM interaction strengths [by means of different values of negative $\theta_{CW}$ at the different applied magnetic fields, as revealed by the respective Curie-Weiss fittings, shown in Fig. 5(a)] is also evident in SLIO, likely in consistent with many other existing QSL systems~\cite{nagprl1,nagprbdp,bcio-ownprb,nagjmmm}.

\subsection{Heat capacity}
The magnetic frustration of any disordered material is commonly expressed by the amount of magnetic entropy retained within the system at very low temperatures. To further check the magnetic ground state and also in order to probe the nature of magnetic excitations in this $d^4$ columnar iridate SLIO, the heat capacity $C_p$ was measured as a function of temperature in both zero field and several higher magnetic fields. The collected $C_p$ versus $T$ data (Fig. 5(d)) do not show any sharp $\lambda$-like anomaly, hallmark of thermodynamic phase transition into a long-range magnetically ordered state, in any of the applied magnetic fields in the measured temperature range, further affirming the absence of magnetic long-range ordered state and/or structural phase transition in SLIO. But, a weak hump-like broad feature develops below $\sim$ 10 K which gradually shifts towards higher temperatures with the increase of applied magnetic fields. As attempt to model this low-temperature weakly field-dependent $C_p$($T$) behavior with the ``two-level Schottky anomaly" description fails, the possible presence of isolated paramagnetic centers/impurities, often highlighted in describing finite magnetism of the otherwise nonmagnetic $d^4$ iridates~\cite{antisite5+prb,byio-asd,psmio-49,cdir2-38,sniochap-46}, could definitely be discarded in the SLIO \cite{4,cdir2-38,chap7-28}. Hence, the total $C_p$ of SLIO can be modeled by the sum of the lattice part ($C_{lattice}$), and the magnetic contribution ($C_m$) from the correlated Ir$^{5+}$ magnetic moments.

The $C_{lattice}$ was obtained after fitting the high temperature $C_p$ data (100-300 K) using Debye-Einstein model, yielding a Debye temperature $\Theta_D$ $\approx$ 425 K. This fit is thereafter extrapolated down to the lowest measuring temperature (see inset to Fig. 5(d)) and taken as the $C_{lattice}$, which was then subtracted from the total $C_p$, so that the sample is now only left out with correlated magnetic contributions to the heat capacity due to intrinsic Ir$^{5+}$ local moments.
\par
Finally, the temperature dependence of the magnetic specific heat $C_m$ is displayed in Fig. 5(e), which reveals a noticeable field-dependence in the 2-12 K temperature range, indicating presence of short-range magnetic correlations in SLIO. As displayed in Fig. 5(e), the $C_m$ attains a strictly field-independent broad maximum at $\sim$ 20-25 K region, suggesting highly frustrated nature of magnetic interactions in SLIO system, similar to the observations in spin-liquids \cite{3,chap7-19,sniochap-64}. In addition, the $C_m$ shows a finite $T$-linear contribution ($C_M$ $\approx$ $\gamma$$T$ at very low temperatures, shown in the inset to Fig. 5(e)), unusual for charge insulators, thus signifying low-energy gapless spin excitations or the presence of metal-like spinon Fermi surface, as discussed in the context of QSL candidates \cite{3,nagprl1,nagprbdp,sniochap-65,sniochap-66,sniochap-67,sniochap-68}.
\par
The release in zero field magnetic entropy $S_M$ (Fig. 5(f)), obtained by integrating the zero field $\frac{C_M}{T}$ with $T$, achieves a value of $\sim$ 2.0 J/mol-K till 70 K, which is almost $\approx$ 20\% of the maximum of $R$$\ln$(2$J$ + 1), with $J$ = 1 corresponding to the hypothetical triplet state~\cite{bcio-ownprb}, for a completely magnetically ordered state. This points towards the fact that despite having large local Ir$^{5+}$ moments as well as significantly strong AFM correlation among these moments, the SLIO preserves a large percentage of the entropy ($\approx$ 80\%), and hence, affirming persistence of spin-fluctuation and also low-energy spin-excitations in SLIO~\cite{6,snio-35,snio-36,snio-38}, thus promoting a potential QSL candidature in SLIO.

Ultimately by comparing the bandwidth (theoretically predicted) {\it vs.} moment (obtained from DFT calculations and experimental observations, either) trends in SLIO with the other two columnar counterparts Sr$_3$NaIrO$_6$ (SNIO) and Sr$_3$KIrO$_6$ (SKIO), and a 6$H$ hexagonal perovskite iridate BZIO \cite{arxiv-snio}, we infer that the magnetic moment continues to decrease with the gradually increasing bandwidth, which can only be argued within the framework of weak or at most intermediate spin-orbit coupling for the $d^4$ iridates, contrary to the pertaining assumptions of strong SOC in iridates.

\section{Summary and Conclusion}
We investigate a new 2$H$-prototype columnar spin chain $d^4$ iridate Sr$_3$LiIrO$_6$ through combined structural, chemical, electronic, transport, magnetic and thermodynamic characterizations. Our dc susceptibility together with the spin polarized electronic structure calculations refutes the coveted nonmagnetic $J_{eff}$=0 singlet ground state in this stoichiometrically perfect structurally ordered material, and rather a large magnetic moment is developed at the individual Ir$^{5+}$-site of SLIO likely due to intersite hopping (via extended Ir-O-O-Ir superexchange pathways within intra- and inter-chains), noncubic crystal distortion, hybridization of Ir-5$d$ states, Ir-O covalent interactions and bandwidth effect. Further, the substantial intra- and inter-chain hopping interaction strengths establish an effective 3-dimensional magnetic environment. Despite significant AFM interactions between the large local Ir$^{5+}$ moments, the heat capacity and dc magnetic susceptibility measurements clearly discern the absence of magnetic ordering in SLIO down to at least 2 K as a result of persistent spin-fluctuation due to the geometric frustration, hence proposing this material as a potential QSL candidate. Both the XPS valence band and electrical resistivity measurements suggest a gapped electronic structure of this compound, but the linear behavior of the magnetic specific heat at very low temperatures points towards the gapless nature of spin-excitations ({\it i.e.}, gapless spin density of states) in this material. 

\section{Acknowledgement}
A.B. thanks CSIR, India and IACS for supporting fellowship. A.B. also acknowledges SERB, DST, India for National Postdoctoral Fellowship (N-PDF, File NO. PDF/2020/000785). S.R. acknowledges the Department of Science and Technology (DST) [Project No. WTI/2K15/74] for support. AB and SR thank Jawaharlal Nehru Centre for Advanced Scientific Research from DST-Synchrotron-Neutron project, for performing experiments at ESRF (Proposal No. HC-2872) synchrotron radiation facility, Grenoble, France. AB and SR thank ISIS muon facility, UK, for performing the muon-spin-relaxation measurements. The authors
also thank the Indian Institute of Technology, Bombay, and the Indian Association for the Cultivation of Science, Kolkata, for support in research.

\end{document}